\PassOptionsToPackage{pdftex}{graphics}
\PassOptionsToPackage{pdftex}{graphicx}
\documentclass[journal]{IEEEtran}
\usepackage{amsmath,amsfonts,amssymb}
\usepackage{orcidlink}
\usepackage{acronym}

\usepackage{xcolor}
\usepackage{spotcolor}

\usepackage{tikz}
\usetikzlibrary{
    calc,
    matrix,
    positioning,
    shapes.multipart,
    fit,
    shapes.geometric,
    3d,
    arrows.meta
}
\tikzset{
  every node/.append style = {font=\footnotesize}
}

\begin{document}

\title{5G Positioning Reference Signal impact assessment in Non-Terrestrial Networks communication service}

\author{Alejandro~Gonzalez-Garrido$^{\dagger}$, Ottavio~M.~Picchi$^{\ddagger}$, and Francesco~Menzione$^{\dagger}$\\%
\emph{$^{\dagger}$European Commission, Joint Research Centre, Ispra, Italy.}\\%
\emph{$^{\ddagger}$External consultant, European Commission.}%
}

\maketitle

\begin{abstract}

5G new radio (NR) non-terrestrial network (NTN) extends cellular connectivity through low Earth orbit (LEO)/ medium Earth orbit (MEO) satellite constellations (among other scenarios) and enables the reuse of downlink (NR) positioning reference signal (PRS) to deliver global positioning, navigation and timing (PNT) capabilities alongside broadband services. However, NTN geometry introduces large inter-satellite differential propagation delays (up to milliseconds), which can cause PRS bursts from non-serving satellites to overlap with the serving-satellite data stream. This paper assesses such coexistence by deriving a tractable statistical model for the slant-range distribution over the visible spherical cap and extending it to a dual-shell constellation through a mixture formulation, yielding a closed-form cumulative density function (CDF) for the differential delay. The model is validated against a 10-day orbit simulation representative of a dual-shell European (NTN) concept. The paper further presents the detection limits of the (PRS) from non-serving satellites under interference from the serving-satellite data stream, expressed in terms of the effective carrier-to-noise density ratio. Finally, the impact of periodic (PRS) transmissions on the uncoded bit error rate (BER) is quantified using standardized (NR) configurations representative of frequency region 1 (FR1) and frequency region 2 (FR2). Monte Carlo results show that the probability of simultaneous multi-PRS overlap during a PRS occasion is low (sub-percent to a few percent, depending on PRS duration and repetition period), while PRS detection remains feasible in the presence of interference from the data stream. Moreover, when the PRS is received at a sufficiently low power (e.g., $\approx 25$~dB below the data signal), the induced degradation on uncoded BER is negligible across a wide range of PRS repetition periods. Conversely, observable BER penalties scale with the PRS duty cycle. These results indicate that NR-PRS-based PNT in NTN is compatible with broadband downlink under suitable periodicity design, motivating further investigation of tracking-loop robustness, coded link performance (e.g., low density parity check (LDPC)), user terminal antenna, and system-level PRS scheduling/orchestration.

\end{abstract}


\section{Introduction}

\Ac{5G} \ac{NR} \ac{NTN} extends connectivity beyond terrestrial infrastructure by integrating satellite links, particularly \ac{LEO} and \ac{MEO} constellations, into the \ac{3GPP} ecosystem~\cite{saad_non-terrestrial_2024}. Beyond broadband access, \ac{NTN} create an opportunity to also provide global \ac{PNT} services~\cite{cianca_leo-based_2024,m_picchi_fused_2025}. The \ac{PNT} service can be supplied by reusing the terrestrial \ac{NR} \ac{PRS} within the same downlink, targeting coexistence between services~\cite{peral-rosado_positioning-enabled_2024}.

A requirement in \ac{5G}-\ac{NTN} is the use of \ac{GNSS} receivers at the \ac{UE} to pre-compensate and assist the initial access to the \ac{NTN} network (TS~38.300, \emph{NR and NG-RAN Overall description; Stage-2}). The trend, however, is to move towards a network independent of third-party systems such as \ac{GNSS}~\cite{gallardo-duval_towards_2026}. To enable such an independent source of \ac{PNT}, the \ac{PRS} needs to be transmitted periodically, and the open question is whether the resulting interference on the data service is affordable, i.e.\ whether the data service can stay within the expected performance bounds.

In the terrestrial scenario, the \ac{5G} \ac{PNT} services physical layer is designed in such a way that the \ac{PRS} maximum differential delay stays within the size of the cell (up to 150~km in some cases)~\cite{dwivedi_positioning_2021}. This implies that the time difference of arrival of two \ac{PRS} is smaller than a \ac{PRS} occasion, so that they overlap only during the \ac{PRS} occasions.

In the \ac{NTN} scenario, the \ac{PNT} services generalizes to single-epoch multi-satellite ranging, conceptually analogous to \ac{GNSS} but using \ac{NR} waveforms~\cite{peral-rosado_positioning-enabled_2024}. In the considered scenario, a serving satellite illuminates a ground cell as part of the communication service, while at least three additional satellites with overlapping coverage can be instructed to transmit \ac{PRS} during predefined slots called \emph{\ac{PRS} occasions}~\cite{xv_joint_2024}. Satellites are assumed time-synchronized to a common reference such that \ac{PRS} occasions are transmitted simultaneously in the satellite time frame. The differential propagation distances cause these signals to arrive at the \ac{UE} at different times. The fact that the \ac{NTN} differential delay exceeds a typical \ac{5G} slot, and therefore that \ac{PRS} bursts from non-serving satellites can overlap with subsequent data slots of the serving satellite, has been noted previously in the literature~\cite{grec_direct--device_2025}. Recent \ac{NTN} works~\cite{gonzalez-garrido_interference_2024} model the interference of bursty \ac{PRS} transmissions, showing that the aggregate \ac{PRS} interference (up to four transmitters) exhibits a heavy-tailed behaviour described by a \ac{GEV} distribution. Under such conditions, treating the \ac{PRS} interference as Gaussian yields a conservative worst-case bound on achievable data rate, whereas a \ac{GEV} model can capture the statistics more accurately and improve the achievable data rate estimation.

The present contribution complements these observations by providing (i) a closed-form geometrical statistical model of the differential delay over a dual-shell constellation, (ii) a detection-limit analysis expressed in terms of the effective $C/N_0$, and (iii) a quantitative assessment of the uncoded \ac{BER} impact on the data link as a function of \ac{PRS} duty cycle and carrier-to-interference ratio.

\section{System models}

This section first describes the geometrical model used to evaluate the probability of interference between different satellite transmissions, and then introduces the signal model and the standardized configurations used in the simulator for the analysis of the interference effect on the data service.

\begin{figure}
    \centering
    \begin{tikzpicture}[scale=0.85, every node/.style={font=\footnotesize}]
% Draw the sphere where the satellites are
\shade[ball color=white, opacity=0.2] (0,0) circle (3cm);
\shade[ball color=white, opacity=0.05] (0,0) circle (4.5cm);
% Draw the Earth
\shade[ball color=blue, opacity=0.3] (0,0) circle (1.5cm);

% Earth centre O
\fill (0,0) circle (1pt);
\node[below right, inner sep=1pt] at (0,0) {$O$};

% User radial (extended, dotted) -- visual reference for Psi
\draw[densely dotted, gray!80] (0,0) -- (0, 3.3);

% Re
\draw[<->] (0,0) -- (-1.5, 0) node[midway, below] {$R_E$};
% h1
\draw[<->] (-1.5,0) -- (-3, 0) node[midway, below] {$h_{\text{1}}$};
\draw[<->,  rotate around={20:(0,0)}] (-1.5,0) -- (-4.5, 0) node[midway, below] {$h_{\text{2}}$};

% Typical user U
\node at (0,1.5) [circle,fill=red!50,inner sep=1.5pt] {};
\node[anchor=south west, inner sep=2pt] at (0,1.5) {$U$};

% Cap base (local horizon LH)
\draw[dashed, red] (2.55,1.5) arc[start angle=0, end angle=360, x radius=2.55cm, y radius=0.5cm] node[midway, below, black, xshift=-0.5cm] {LH};

% Slant max at horizon
\draw[->] (0,1.5) -- (2.55, 1.5) node[midway, below] {$R_{\text{MAX}}$};

% Cap base for theta mask
\draw[dashed, green] (2.2,2) arc[start angle=0, end angle=360, x radius=2.2cm, y radius=0.3cm] node[midway, above, black, xshift=-0.5cm] {$\theta_{\text{MASK}}$};

% Spherical cap (shell 1)
\draw[dashed, green] (2.2,2) arc[start angle=43, end angle=135, x radius=3cm, y radius=3cm];

% ------------------- Blue spherical cap (larger) -------------------
\def\rBaseBlueX{4cm}
\def\rBaseBlueY{0.5cm}
\def\rCapBlue{6cm}
\draw[dashed, blue] (4,2)
  arc[start angle=0, end angle=360, x radius=\rBaseBlueX, y radius=\rBaseBlueY];
\draw[dashed, blue] (4,2)
  arc[start angle=43, end angle=135, x radius=5.5cm, y radius=8cm];

% Slant max at theta mask (rho_{MAX,1})
\draw[->] (0,1.5) -- (2.2, 2) node[right] {$\rho_{\text{MAX},1}$};

% =================================================================
% Angle annotations on a representative shell-1 satellite S
% =================================================================
\coordinate (Sill) at (-1.5, 2.5);

% Radial line from O to S (dotted)
\draw[densely dotted, gray!80] (0,0) -- (Sill);

% Slant range rho from U to S
\draw[->, thick] (0,1.5) -- (Sill);
\node at (-0.55, 2.12) {$\rho$};

% Satellite S marker (on top of the lines) and label
\node at (Sill) [star,fill=green, scale=0.5] {};
\node[anchor=south, inner sep=1pt] at (-1.5, 2.62) {$S$};

% Psi: geocentric angle at O between user-radial (90 deg) and OS direction (~121 deg)
\draw[->] (0, 0.85) arc[start angle=90, end angle=121, radius=0.85cm];
\node at (-0.30, 1.05) {$\Psi$};

% e: elevation at U between local horizon (180 deg, leftward) and slant range US (~146 deg)
\draw[->] (-0.55,1.5) arc[start angle=180, end angle=146.3, radius=0.55cm];
\node at (-0.78, 1.78) {$e$};

% phi: azimuth indicator on the local horizontal plane (small flat arc near U)
\draw[->, gray!70] (0.65, 1.5) arc[start angle=0, end angle=170, x radius=0.65cm, y radius=0.13cm];
\node[below, gray!60!black, inner sep=1pt] at (0.30, 1.42) {$\phi$};
% =================================================================

% Other satellites
\node at (1.5,4) [star,fill=blue, scale=0.5] {};
\node at (-3,3) [star,fill=blue,scale=0.5] {};
\node at (-0.8,2.7) [star,fill=green, scale=0.5] {};
\node at (0.8,2.2) [star,fill=green,scale=0.45] {};
\node at (2,-2) [star,fill=black,scale=0.5] {};

% Labels
\node at (2.2, 2.8) {Cap $A$};
\node at (3, -1.5) [circle,fill=red!50,scale=0.5]{};
\node at (3.3, -1.5) [right]{Ground receiver};
\node at (3, -2) [star,fill=green,scale=0.5]{};
\node at (3.3, -2) [right]{Satellites in shell 1};
\node at (3, -2.5) [star,fill=blue,scale=0.5]{};
\node at (3.3, -2.5) [right]{Satellites in shell 2};
\node at (3, -3) [star,fill=black,scale=0.5]{};
\node at (3.3, -3) [right]{Not visible satellite};
\end{tikzpicture}
    \caption{Geometry of the analytical model. The user sees satellites in two shells; the visible spherical cap is bounded by the minimum elevation $e_{\min}=\theta_{\mathrm{MASK}}$, defining a maximum slant range $\rho_{\max,1}$ on shell~1. The relevant angles are the geocentric angle $\Psi$ at the Earth centre between the user radial and the satellite radial, the elevation angle $e$ at the user, and the azimuth $\phi$ around the local vertical; for shell~1 the corresponding limits are $\Psi_{\max,1}$ and $\rho_{\max,1}$.}
    \label{fig:sat-DoC}
\end{figure}
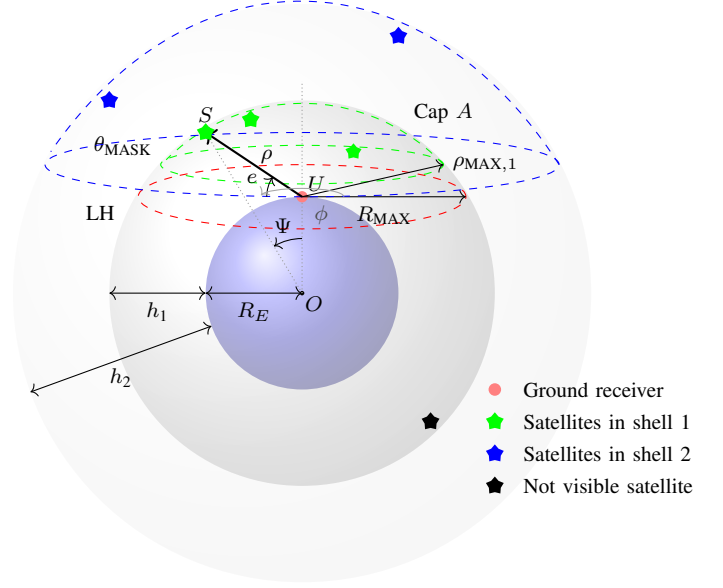

This work considers a downlink \ac{NR}-\ac{NTN} deployment where each satellite hosts a \ac{SAN} (\ac{gNB}-equivalent) illuminating the Earth with spot beams. A generic \ac{UE} located in a ground cell is served on the downlink by one ``serving'' satellite and may receive additional \ac{PRS} transmissions from one or more non-serving satellites in \ac{LoS}, as depicted in Fig.~\ref{fig:sat-DoC}. We assume that the network orchestrator periodically schedules a subset of nearby non-serving satellites to illuminate the same cell as the serving satellite during predefined \ac{PRS} occasions. In other words, the same ground footprint is illuminated by multiple satellites for \ac{PNT} purposes. The associated resource cost on the multiple satellites is the price for an independent \ac{PNT} service, and the analysis in this paper is intended to inform whether that cost is acceptable from a data-link standpoint. Detailed multi-satellite beam scheduling and orchestration is outside the scope of this paper.

A specific impairment of the \ac{NTN} scenario is the large differential propagation delay between satellites observed at different elevation angles and in different shells, as illustrated in Fig.~\ref{fig:sat-DoC}. Using the \ac{3GPP} reference scenarios, one-way delay differentials between a satellite at the horizon and one overhead reach up to $\sim 3.18$~ms for \ac{LEO} and above 10~ms for \ac{MEO}. These values exceed the \ac{NR} slot duration across the main numerologies used in \ac{NTN} (e.g., 0.5~ms at 30~kHz \ac{SCS} or 0.125~ms at 120~kHz \ac{SCS}). Consequently, even if \ac{PRS} transmissions are time-aligned at the satellite reference, the received \ac{PRS} at the \ac{UE} spans multiple downlink slots and may overlap with subsequent data slots, producing cross-service interference at the \ac{UE}~\cite{grec_direct--device_2025}. The \ac{PRS} from a non-serving satellite therefore collides with the data service of the serving satellite, creating a burst of interference whose duration is the number of \ac{PRS} symbols per occasion (per interfering satellite).

\subsection{Geometrical modelling}

In \ac{NTN}, differential delays across a beam/cell can be on the order of hundreds of microseconds to milliseconds, depending on orbit and reference point definition. \ac{3GPP} explicitly discusses such values (e.g., \ac{NGSO} examples and maximum differential delay) and the need for compensation~\cite{3gpp_38821_2021}. This subsection develops a statistical model for these differential delays leading to a closed-form expression for the \ac{CDF}. The model relies on the following assumptions, which are summarised here for clarity:
\begin{itemize}
\item the Earth is a sphere of radius $R$;
\item each satellite shell is a spherical shell at altitude $h_i$, so that the orbital radius is $a_i=R+h_i$;
\item conditional on the \ac{LoS} constraint $e\geq e_{\min}$, the direction from the \ac{UE} to a randomly chosen visible satellite is uniformly distributed over the visible spherical cap. This is a tractable approximation: in real constellations with high-inclination orbits, the local satellite density depends on \ac{UE} latitude, and the actual distribution may differ. The validation in Section~III shows that the closed-form \ac{CDF} reproduces the shape of the simulated distribution well enough for the present interference analysis;
\item the two contributing satellites are drawn independently.
\end{itemize}

\subsubsection{Slant range PDF for a single satellite}
Let $\Psi\in[0,\pi]$ be the geocentric angle between the user's radial vector and the satellite's radial vector, $\phi\in[0,2\pi)$ the local azimuth around the user's vertical (i.e.\ the angle that, together with $\Psi$, parameterises the satellite direction over the visible spherical cap), and $e$ the local elevation of the satellite seen from the user. Then the slant range is
\begin{equation}
    \rho(\Psi)=\sqrt{R^2+a^2-2Ra\cos{\Psi}},
    \label{eq:slant_range}
\end{equation}
and the elevation at the user satisfies
\begin{equation}
    \sin{e}=\frac{a\cos{\Psi}-R}{\rho(\Psi)}.
    \label{eq:elevation}
\end{equation}

The \ac{LoS} elevation mask $e\geq e_{\min}$ implies $\Psi\leq\Psi_{\max}$, where $\Psi_{\max}$ solves~\eqref{eq:elevation} with $e=e_{\min}$ and can be written as
\begin{equation}
    \cos{\Psi_{\max}}=\frac{R\cos^2{e_{\min}}+\sin{e_{\min}}\sqrt{a^2-R^2\cos^2{e_{\min}}}}{a}.
\end{equation}

The maximum admissible slant range is $\rho_{\max}=\rho(\Psi_{\max})$ and the minimum $\rho_{\min}=\rho(0)=h$.

Under the assumption that, conditional on visibility, $(\Psi,\phi)$ is uniformly distributed over the visible spherical cap, the marginal density of $\Psi$ is
\begin{equation}
    f_{\Psi}(\psi)=\frac{\sin{\psi}}{1-\cos{\psi_{\max}}},\quad 0\leq\psi\leq\psi_{\max}.
\end{equation}

Defining $d\triangleq \rho_{\max}^{2}-h^2$ and applying~\eqref{eq:slant_range}, the slant range \ac{PDF} and \ac{CDF} for a single satellite become
\begin{equation}
    f_{\rho}(\rho)=\frac{2\rho}{d}, \quad F_{\rho}(\rho)=\frac{\rho^2-h^2}{d}, \quad h\leq\rho\leq\rho_{\max}.
    \label{eq:pdf_slant}
\end{equation}

\subsubsection{Differential slant range for two satellites in two shells}
This work investigates inter-satellite interference for the case of a constellation is composed of two shells, one at $h_1=1200$~km and the other at $h_2=8000$~km, consolidating the works in~\cite{picchi_fused_2025, m_picchi_fused_2025}. Assuming two satellites drawn independently and possibly from different shells, the absolute differential delay becomes a mixture over the four shell combinations.

Let shell $i\in\{1,2\}$ have altitude $h_i$ and orbit radius $a_i=R+h_i$. Under the same uniform-spherical-cap \ac{LoS} model with mask $e_{\min}$, each shell yields a range support $\rho\in[\rho_{\min,i},\rho_{\max,i}]$ with $\rho_{\min,i}=h_i$ and $\rho_{\max,i}=\rho_i(\psi_{\max,i})$, and a range \ac{PDF}/\ac{CDF} as in~\eqref{eq:pdf_slant}.

Let $p\triangleq \Pr[\text{a randomly LoS satellite is from Shell 1}]$. Then the shell pair $(I,J)\in\{1,2\}^2$ has weights $w_{1,1}=p^2$, $w_{2,2}=(1-p)^2$, and $w_{1,2}=w_{2,1}=p(1-p)$. When the number of satellites in one shell is much larger than in the other, $p$ can be approximated by
\begin{equation}
    p=\frac{N_1 q_1}{N_1 q_1+N_2 q_2},
\end{equation}
where $N_i$ is the number of satellites in shell $i$ and $q_i=\tfrac{1-\cos\psi_{\max,i}}{2}$ is the fractional area of the visible cap on shell $i$.

For a fixed ordered pair $(i,j)$ and an absolute differential range $x\geq 0$, the \ac{CDF} of $|\rho_i-\rho_j|$ is
\begin{equation}
    F_{|\Delta\rho|\,|\,i,j}(x) = \!\!\int_{\rho_{\min,j}}^{\rho_{\max,j}}\!\!\! f_{\rho_j}(u)\bigl[F_{\rho_i}(u+x)-F_{\rho_i}(u-x)\bigr]\,du,
    \label{eq:cdf_abs_dr}
\end{equation}
where $F_{\rho_i}(\cdot)$ is taken as zero below $\rho_{\min,i}$ and as one above $\rho_{\max,i}$. The mixture \ac{CDF} for the absolute differential delay is then
\begin{equation}
    F_{|\Delta\tau|}(t)=\Pr[|\Delta\tau|\leq t]=\sum_{i,j}w_{i,j}\,F_{|\Delta\rho|\,|\,i,j}(ct), \quad t\geq 0.
    \label{eq:cdf_abs_dtau}
\end{equation}

\subsection{Signal model}

\subsubsection{Waveform model}
The waveform used in \ac{5G} \ac{NTN} for the communication service is \ac{CP}-\ac{OFDM} as defined in \ac{3GPP} TS~38.211 Section~5.3.1, with bandwidth $BW$~Hz divided into $N_{\mathrm{scs}}$ subcarriers. The subcarrier spacing is $f_{\mathrm{scs}}=1/T_{\mathrm{u}}$ and the symbol duration is $T_{\mathrm{s}}=T_{\mathrm{u}}+T_{\mathrm{cp}}$, where $T_{\mathrm{cp}}$ is the duration of the \ac{CP}.

Each \ac{OFDM} slot is composed of 14 symbols, and each \ac{OFDM} subframe contains a different number of slots depending on the \ac{5G} numerology. The subframe duration is fixed to 1~ms. The \ac{RG} is the matrix $\mathbf{A}\in\mathbb{C}^{M\times N_{\mathrm{scs}}}$ whose entries $A[m,k]$ (\acp{RE}) carry constellation symbols, control information, or pilots, as configured by upper layers. In the simulator used in Section~III, user and control \acp{RE} carry random \ac{QPSK} symbols, while the pilot \acp{RE} carry the \ac{PRS} sequence defined in \ac{3GPP} TS~38.211 Section~7.4.1.7.3~\cite{3gpp_nr_2024}.

The baseband transmitted signal is modelled as
\begin{equation}
    x(t)=\sqrt{P_{\mathrm{TX}}}\!\!\sum_{m=0}^{M-1}\sum_{k=0}^{N_{\mathrm{scs}}-1}\!\!A[m,k]\,e^{j2\pi k f_{\mathrm{scs}}(t-mT_{\mathrm{s}})}\,g\!\left(t-mT_{\mathrm{s}}\right),
    \label{eq:txsignal}
\end{equation}
where the rectangular pulse shape
\begin{equation*}
    g(t)=
    \begin{cases}
    1,& -T_{\mathrm{cp}}<t<T_{\mathrm{u}},\\
    0,& \text{otherwise,}
    \end{cases}
\end{equation*}
preserves subcarrier orthogonality while keeping the simulator simple. $P_{\mathrm{TX}}$ is the transmit power and is fixed at the satellite \ac{HPA} to guarantee a minimum performance for users at the beam edge.

\subsubsection{Received signal model}

The \ac{UE} receives the superposition of all \ac{LoS} \ac{SAN} signals, each with its own delay, Doppler, and complex channel coefficient. In continuous time,
\begin{equation}
    r(t)=\sum_{i=0}^{S-1} \gamma_i\, x_i\!\left(t-\tau_i\right) e^{j(2\pi \nu_i t+\theta_i)} + w(t),
    \label{eq:rxsignal}
\end{equation}
where $\gamma_i\in\mathbb{C}$ captures the magnitude (path loss, satellite \ac{EIRP}, \ac{UE} antenna gain in the direction of satellite $i$, beam pointing loss) and the random carrier phase $\theta_i$, $\tau_i$ and $\nu_i$ are the propagation delay and Doppler shift of satellite $i$, and $w(t)$ is \ac{AWGN}. The serving satellite is indexed $i=0$. Equivalently, we will write $|\gamma_i|^2=P_{\mathrm{RX},i}$ and reserve the symbol $C\triangleq P_{\mathrm{RX},0}$ for the received carrier power from the serving satellite. The received carrier power from a non-serving satellite is therefore $|\gamma_i|^2=P_{\mathrm{TX}}\,G_{\mathrm{sat},i}\,G_{\mathrm{UE}}(\Omega_i)\,/\,L_{\mathrm{fs},i}$, where $L_{\mathrm{fs},i}$ is the free-space loss for slant range $\rho_i$, $G_{\mathrm{sat},i}$ is the satellite antenna gain toward the user, and $G_{\mathrm{UE}}(\Omega_i)$ is the \ac{UE} antenna gain in the direction $\Omega_i$ of satellite $i$. The latter captures the spatial filtering provided by the \ac{UE} antenna pattern; in \ac{FR2} configurations using directive arrays~\cite{andres_vasquez-peralvo_multibeam_2024}, $G_{\mathrm{UE}}(\Omega_i)$ for $i\neq 0$ is well below the main-beam gain.

\subsection{Interference and BER model}
\label{sec:interference_model}

A distinctive feature of the \ac{NTN} scenario is that the inter-satellite relative delay can be orders of magnitude larger than the \ac{CP}, producing \ac{ISI}/\ac{ICI} when signals overlap in time and frequency. Once the \ac{CDF} of $|\Delta\tau|$ in~\eqref{eq:cdf_abs_dtau} is known, an envelope on the per-event interference power is obtained as a transformation of $\Delta\tau$, taking into account the relative \ac{FSPL} between the serving and non-serving slant ranges and the  antenna gains.

We define the \ac{PRS} \emph{duty cycle} as the fraction of time during which the \ac{PRS} is active on the resource grid of the serving link,
\begin{equation}
    D \triangleq \frac{T_{\mathrm{PRS}}}{T_{\mathrm{rep}}}\in[0,1],
\end{equation}
where $T_{\mathrm{PRS}}$ is the per-occasion \ac{PRS} burst duration in \ac{OFDM} symbols (converted to seconds) and $T_{\mathrm{rep}}$ is the \ac{PRS} \emph{repetition period}, i.e.\ the time between consecutive \ac{PRS} occasions. The two parameters are distinct: $T_{\mathrm{rep}}$ controls how frequently \ac{PRS} bursts occur, while $T_{\mathrm{PRS}}$ controls how long each burst lasts.

Assuming both satellites transmit with the same \ac{HPA} setting and treating, for analytical tractability, a single dominant non-serving interferer, the worst-case per-symbol interference power at the \ac{UE} can be approximated as
\begin{equation}
    I(\Delta\tau)=C\,g\,\frac{L_{\mathrm{fs},0}}{L_{\mathrm{fs},1}}=C\,g\left(\frac{\rho_0}{\rho_0+c\,\Delta\tau}\right)^2,
    \label{eq:interference_fspl}
\end{equation}
where $g=G_{\mathrm{UE}}(\Omega_1)/G_{\mathrm{UE}}(\Omega_0)\leq 1$ models the spatial filtering of the \ac{UE} antenna toward the non-serving satellite. Equation~\eqref{eq:interference_fspl} explicitly ignores the Doppler offset $\Delta\nu_i$, the resource-grid structure of $\mathbf{A}$, and the orthogonality preserved by the \ac{CP} when it covers the differential delay; it should be interpreted as an envelope on the worst-case interference power per overlapping \ac{OFDM} symbol, useful to derive analytical sizing rules. The Monte Carlo simulator in Section~III does not rely on~\eqref{eq:interference_fspl}: it uses the full sampled signal model~\eqref{eq:rxsignal} and therefore captures the \ac{ICI} produced by $\Delta\nu_i$ as well as the partial overlap pattern.

With multiple non-serving satellites $\mathcal{I}=\{i:i\neq 0,\,i\in\text{LoS}\}$, the aggregate interference is the sum of the per-satellite contributions, $I_{\mathrm{agg}}(t)=\sum_{i\in\mathcal{I}}I_i(t)$, and a worst-case bound is obtained by replacing $I(\Delta\tau)$ in the following expressions with $\sum_i I(\Delta\tau_i)$. As reported in~\cite{gonzalez-garrido_interference_2024}, the aggregate exhibits a heavy-tailed distribution; treating $I_{\mathrm{agg}}$ as Gaussian (as done here for tractability) is conservative for the data link.

The instantaneous \ac{SINR} at the receiver is $\gamma(\Delta\tau)=C/(N+I(\Delta\tau))$, and when no \ac{PRS} from a non-serving satellite is received, $\gamma_0=C/N$. Denoting by $P_{b}(\cdot)$ the \ac{BER} function versus \ac{SINR} for the modulation used on the data link, the time-averaged uncoded \ac{BER} is
\begin{equation}
    \overline{P_b}=\underbrace{(1-D)\,P_b(\gamma_0)}_{\text{no PRS overlap}}+\underbrace{D\,\mathbb{E}_{\Delta\tau}\!\left[P_b\!\left(\frac{C}{N+I(\Delta\tau)}\right)\right]}_{\text{PRS overlap},}
    \label{eq:BER_interference}
\end{equation}
where for \ac{QPSK} (the modulation used in our simulator) $P_b(\gamma)=Q(\sqrt{2\gamma})$. Equation~\eqref{eq:BER_interference} confirms that the impact of \ac{PRS} interference grows with $D$: a smaller repetition period $T_{\mathrm{rep}}$ (more frequent \ac{PRS} occasions, larger $D$) places more weight on the second term and thus increases the average \ac{BER}.

Two important caveats apply to~\eqref{eq:BER_interference}. First, it is a strictly uncoded analysis: the \ac{3GPP} \ac{NR} physical layer uses \ac{LDPC} block coding, so the burst nature of the interference (concentrated within a few \ac{OFDM} symbols within an \ac{LDPC} codeword) interacts with the channel decoder differently from a uniform \ac{AWGN} channel. A coded-link evaluation, ideally including \ac{LDPC} block error rate as a function of $T_{\mathrm{rep}}$, $T_{\mathrm{PRS}}$, and codeword duration, is part of the future work. Second, modulations beyond \ac{QPSK} (16-/64-/256-\ac{QAM}) are supported in \ac{NR} and would replace $P_b(\gamma)=Q(\sqrt{2\gamma})$ by the appropriate constellation-specific expression.

\section{Simulation and results}

This section presents simulation results for three quantities: the differential delay observed for the constellation under study; the \ac{ROC} of \ac{PRS} detection when the interference is generated by another satellite's \ac{5G} data transmission; and the impact of periodic \ac{PRS} transmissions on the uncoded \ac{BER}. The simulations are carried out for two signal parameter settings, hereafter referred to as \ac{FR1} and \ac{FR2}, respectively. The main simulation parameters are summarised in Table~\ref{tab:simulation_parameters}.

\paragraph*{Antenna and system-scenario assumptions} Unless stated otherwise, the following baseline assumptions are used in all sections of the simulation:
\begin{itemize}
    \item \emph{Satellite}: a single active \ac{PRS} beam per non-serving satellite illuminates the same ground cell as the serving satellite during the \ac{PRS} occasion. The satellite antenna gain toward the user is taken at the maximum of the beam pattern, i.e., \ac{EIRP} per beam is set such that the serving link operates at the configured $C/N_0$ (or equivalently \ac{SNR}); the same per-beam \ac{EIRP} is assumed for the non-serving satellite during the \ac{PRS} occasion.
    \item \emph{\ac{UE} antenna}: in \ac{FR1} a low-directivity antenna with approximately uniform gain over the visible spherical cap is assumed, so that $g\approx 1$ for any non-serving satellite. In \ac{FR2} a directive array~\cite{andres_vasquez-peralvo_multibeam_2024} pointed at the serving satellite is assumed; the worst-case results in this paper use $g$ corresponding to the first sidelobe, which is a representative bound on inter-satellite coupling.
    \item \emph{Beam time evolution}: during a \ac{PRS} occasion both serving and non-serving beams are simultaneously active on the cell; outside the \ac{PRS} occasion, only the serving beam is active. This is the worst case from the data-link perspective.
\end{itemize}

\begin{table}
    \centering
    \caption{Simulation parameters}
    \begin{tabular}{|l|l|l|}
        \hline
        Parameter & Symbol & Value \\
        \hline
        Resource blocks & $R_{\mathrm{B}}$ & 52@FR1 / 92@FR2 \\
        \hline
        Subcarrier spacing & $\Delta f$ & 30~kHz@FR1 / 120~kHz@FR2 \\
        \hline
        Monte Carlo iterations & $M_{\mathrm{c}}$ & 1000 \\
        \hline
        Satellite altitudes & $h$ & 1200~km / 8000~km \\
        \hline
        Number of satellites & $N_{\mathrm{h}}$ & 264@1200~km / 18@8000~km \\
        \hline
        PRS burst duration (default) & $T_{\mathrm{PRS}}$ & 10 \ac{OFDM} symbols\\
        \hline
        Doppler search bin & $\Delta\nu_{\mathrm{bin}}$ & $1/(2T_{\mathrm{PRS}})$ \\
        \hline
    \end{tabular}
    \label{tab:simulation_parameters}
\end{table}

\subsection{Differential delay}

This subsection presents the model for the differential delay where we compare the theoretical \ac{CDF} mixture~\eqref{eq:cdf_abs_dtau} with a realistic simulation of 10 days of satellite passes over a fixed user. The simulation uses the orbital parameters of~\cite{picchi_fused_2025} and samples the satellite-to-user delays every 10~s.

\begin{figure}
    \centering
    \includegraphics[width=1\linewidth]{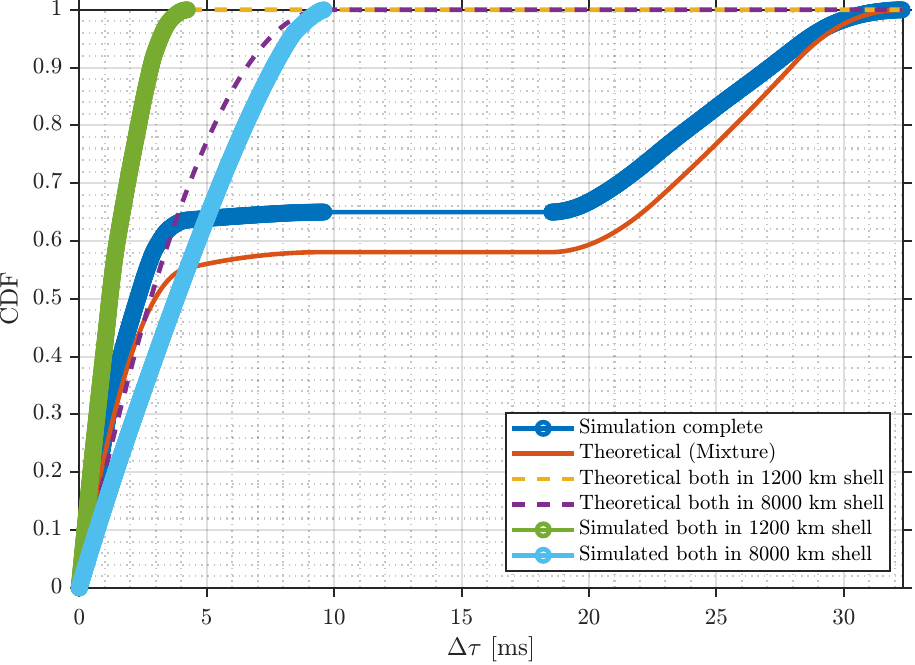}
    \caption{\ac{CDF} of the absolute differential delay $|\Delta\tau|$. The theoretical curves correspond to two satellites both in the 1200~km shell, both in the 8000~km shell, and the dual-shell mixture, and are compared against a 10-day satellite-pass simulation.}
    \label{fig:differential_delay}
\end{figure}

Results in Fig.~\ref{fig:differential_delay} shows a residual mismatch between the theoretical curves and the simulation stems from the uniform-spherical-cap assumption of Section~II.A: the actual angular distribution of satellites depends on the user latitude and on constellation inclination, and is not perfectly uniform. Despite the assumption, the shape of the \ac{CDF} matches, and the qualitative conclusion is robust: the differential delay between two satellites is much larger than one \ac{5G} slot, so that \ac{PRS} transmissions from non-serving satellites systematically collide with the serving-satellite data slots.

Table~\ref{tab:Prob_collision} summarizes the probability that, during a \ac{PRS} occasion, two \ac{PRS} bursts overlap simultaneously at the receiver, as obtained from the simulation. As explained earlier, the \ac{PRS} transmission start times are assumed time-synchronized at the satellite reference. Synchronization errors are neglected, since each satellite is assumed to be equipped with a \ac{GNSS} receiver that disciplines the \ac{5G} transmitter, in line with the multi-tier synchronization paradigm.

\begin{table}
    \centering
    \caption{Probability of simultaneous reception of multiple PRS transmissions}
    \begin{tabular}{|c|c|c|}
        \hline
        $T_{\mathrm{PRS}}$~[\textmu s] & \# Symbols & $\Pr\!\left[|\Delta\tau|\leq T_{\mathrm{PRS}}\right]$ \\
        \hline
        8.33 & 1 & 0.23\% \\
        \hline
        33.32 & 4 & 0.93\% \\
        \hline
        49.98 & 6 & 1.39\% \\
        \hline
        83.33 & 10 & 2.33\% \\
        \hline
    \end{tabular}
    \label{tab:Prob_collision}
\end{table}

\subsection{PRS detection}

We now evaluate the limits on \ac{PRS} acquisition by computing the \ac{ROC} when the interference is the \ac{5G} random data transmission from another (serving) satellite. This represents the case in which a receiver attempts to detect the \ac{PRS} from non-serving satellites while still being illuminated by its serving satellite.

\begin{figure}
    \centering
    \includegraphics[width=1\linewidth]{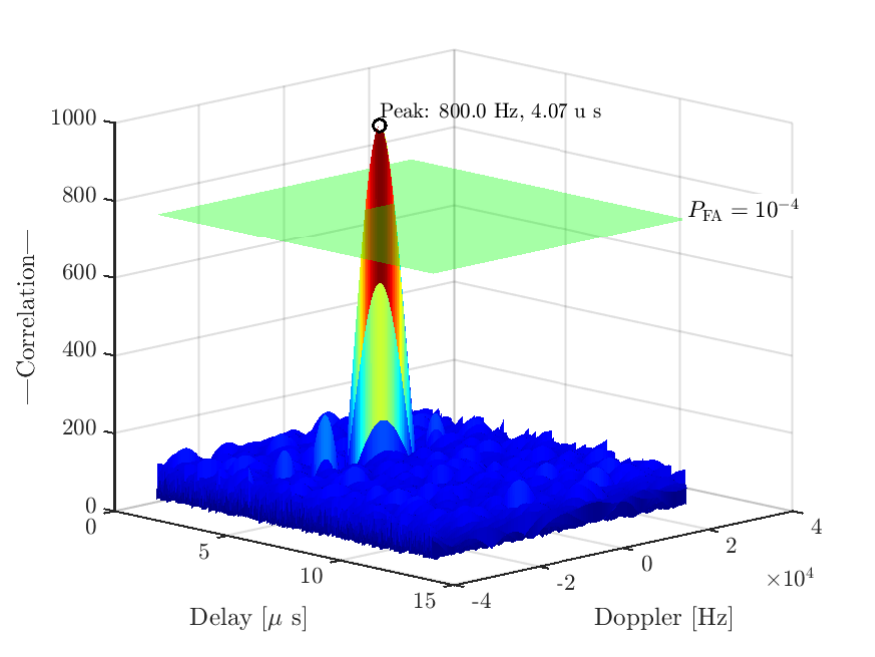}
    \caption{Example of the search grid in delay/Doppler and detection threshold (green) for the \ac{PRS} with a carrier-to-interference ratio $C_{\mathrm{PRS}}/I=-25$~dB relative to the \ac{5G} data signal of the serving satellite.}
    \label{fig:detection}
\end{figure}

The detector performs a two-dimensional search over delay and Doppler, with delay bin equal to $T_u/N_{\mathrm{scs}}$ and Doppler bin $\Delta\nu_{\mathrm{bin}}=1/(2T_{\mathrm{PRS}})$, as the example illustrated in Fig.~\ref{fig:detection}. The Doppler search range is set to $\pm 50$~kHz for \ac{FR1} and $\pm 200$~kHz for \ac{FR2}, accounting for the higher residual Doppler in \ac{FR2}. The contribution of an additional Doppler search dimension to the false-alarm probability is included in the \ac{ROC} curves.

The relevant parameter for the \ac{ROC} is the $C_{\mathrm{PRS}}/N_{0,\mathrm{eff}}$. It is defined as for a target satellite with \ac{PRS} received power $C_{\mathrm{PRS}}$ and aggregate co-band interference power $I$ (other \acp{PRS} and the data signal), the effective carrier-to-noise density is
\begin{equation}
  \left(\frac{C_{\mathrm{PRS}}}{N_0}\right)_{\mathrm{eff}}
  =
  \frac{C_{\mathrm{PRS}}}{N_0+I^{eq}},
  \label{eq:cn0_eff}
\end{equation}
where the aggregate interference is computed as $I^{\mathrm{eq}} = \sum_{i=1}^{S-1} \beta_{0,s} P_{r,i}$, with $\beta_{0,s}$ the \ac{SSC} between the signals.

For the case in which the \ac{UE} has a directive antenna pointing at the serving satellite (\ac{FR2} array), the \ac{PRS}-to-\ac{PRS} interference among non-serving satellites is naturally limited: only one \ac{PRS} component falls within the main beam at a time in most realisations, and the others are attenuated by the antenna pattern. This is captured implicitly through the spatial-filtering factor $g$ in~\eqref{eq:interference_fspl}.

\begin{figure}
    \centering
    \includegraphics[width=1\linewidth]{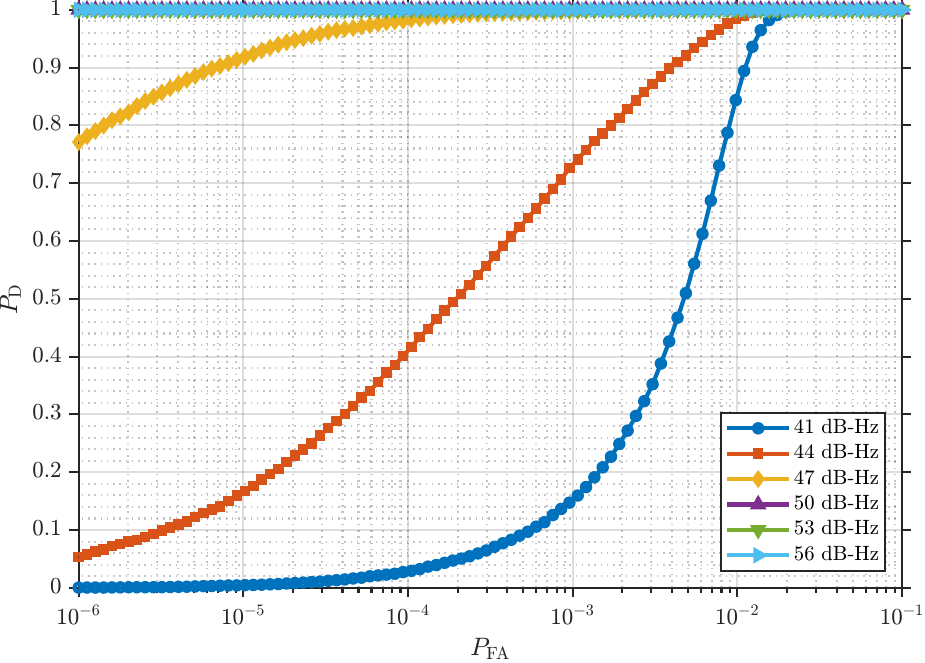}
    \caption{\ac{ROC} of \ac{PRS} detection for different values of $C_{\mathrm{PRS}}/N_{0,\mathrm{eff}}$ of the \ac{PRS} component, with the \ac{5G} data of the serving satellite treated as the dominant interference. \ac{FR1} parameters; $T_{\mathrm{PRS}}=10$ \ac{OFDM} symbols (\textit{see Table~\ref{tab:simulation_parameters}}).}
    \label{fig:ROC_FR1}
\end{figure}

\begin{figure}
    \centering
    \includegraphics[width=1\linewidth]{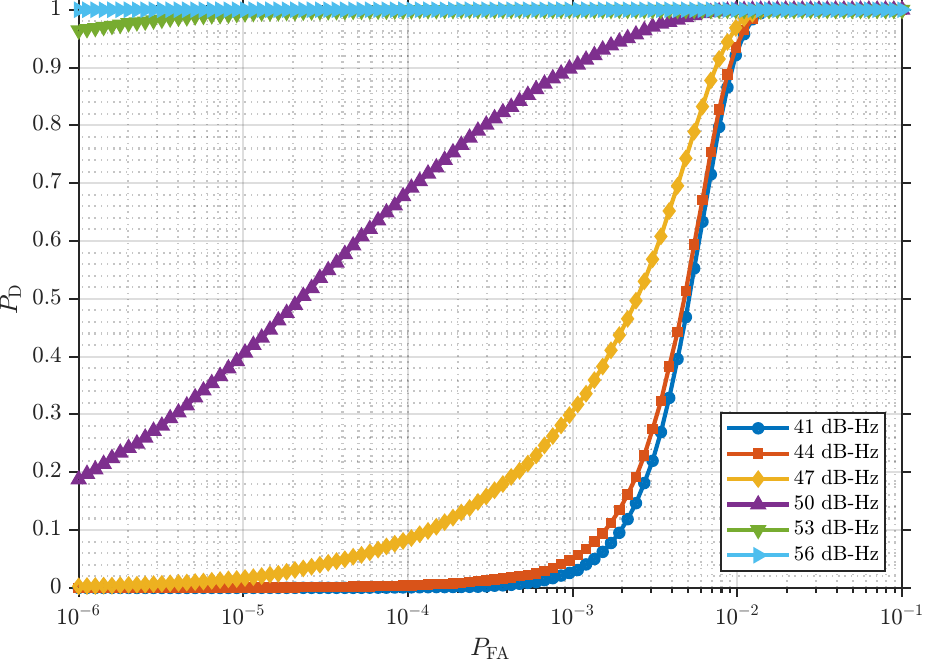}
    \caption{\ac{ROC} of \ac{PRS} detection for different values of $C_{\mathrm{PRS}}/N_{0,\mathrm{eff}}$ of the \ac{PRS} component. \ac{FR2} parameters; $T_{\mathrm{PRS}}=10$ \ac{OFDM} symbols.}
    \label{fig:ROC_FR2}
\end{figure}

The \acp{ROC} in Figs.~\ref{fig:ROC_FR1} and~\ref{fig:ROC_FR2} report the detection probability versus false-alarm probability for several values of the $C_{\mathrm{PRS}}/N_{0,\mathrm{eff}}$ of the \ac{PRS}. They are obtained for a \ac{PRS} burst of $T_{\mathrm{PRS}}=10$ \ac{OFDM} symbols (the longest entry in Table~\ref{tab:Prob_collision}). 

The 6~dB difference between the \ac{FR1} and \ac{FR2} \acp{ROC} reported below is consistent with the ratio of \ac{OFDM} symbol durations at the two configured \acp{SCS} ($T_{\mathrm{sym}}\approx 33.3$~\textmu s at 30~kHz \ac{SCS} versus $T_{\mathrm{sym}}\approx 8.93$~\textmu s at 120~kHz \ac{SCS}, i.e.\ approximately a factor of four shorter, hence 6~dB less integrated energy at the same $C_{\mathrm{PRS}}/N_{0,\mathrm{eff}}$).

\subsection{PRS interference impact on the data service}

\begin{figure}
    \centering
    \includegraphics[width=1\linewidth]{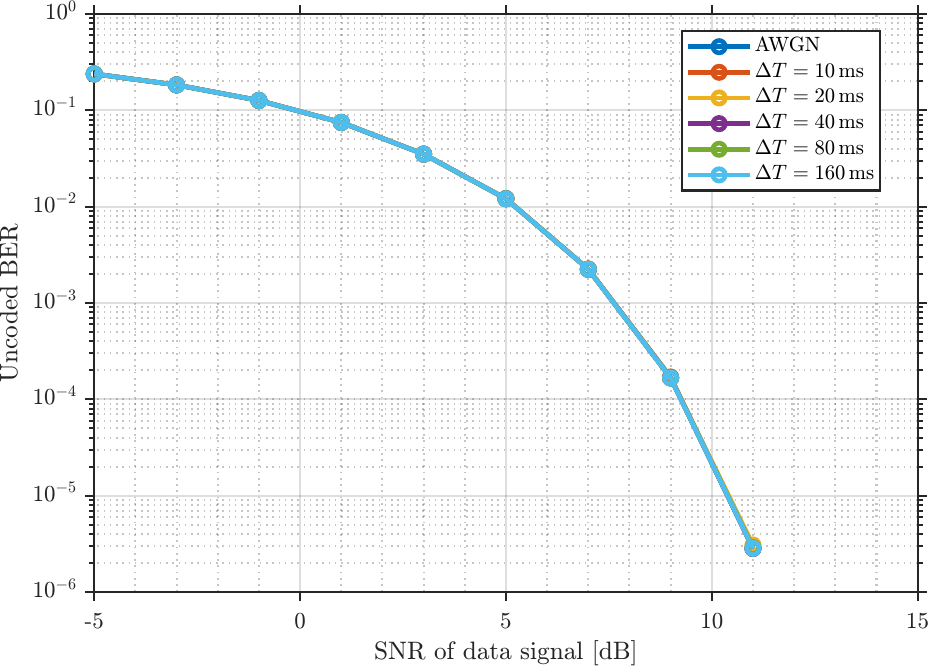}
    \caption{Uncoded \ac{BER} versus data-signal \ac{SNR} for different \ac{PRS} repetition periods $\Delta T=T_{\mathrm{rep}}$. The \ac{PRS} carrier power is 25~dB below that of the \ac{5G} data ($C_{\mathrm{data}}/I=25$~dB). \ac{FR1} parameters.}
    \label{fig:BER_20_FR1}
\end{figure}

\begin{figure}
    \centering
    \includegraphics[width=1\linewidth]{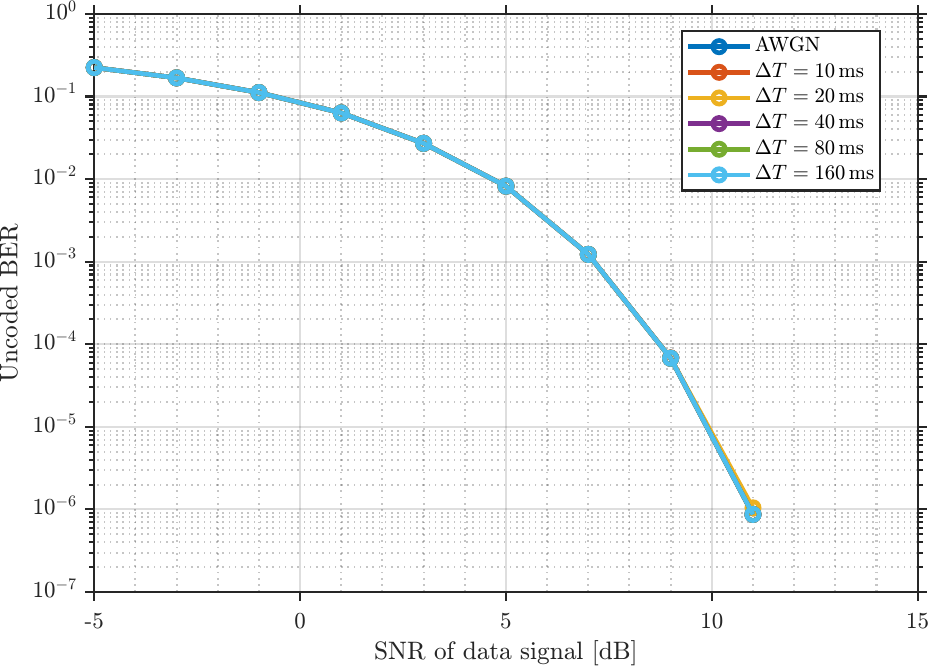}
    \caption{Uncoded \ac{BER} versus data-signal \ac{SNR} for different \ac{PRS} repetition periods $\Delta T=T_{\mathrm{rep}}$. The \ac{PRS} carrier power is 25~dB below that of the \ac{5G} data ($C_{\mathrm{data}}/I=25$~dB). \ac{FR2} parameters.}
    \label{fig:BER_20_FR2}
\end{figure}

\begin{figure}
    \centering
    \includegraphics[width=1\linewidth]{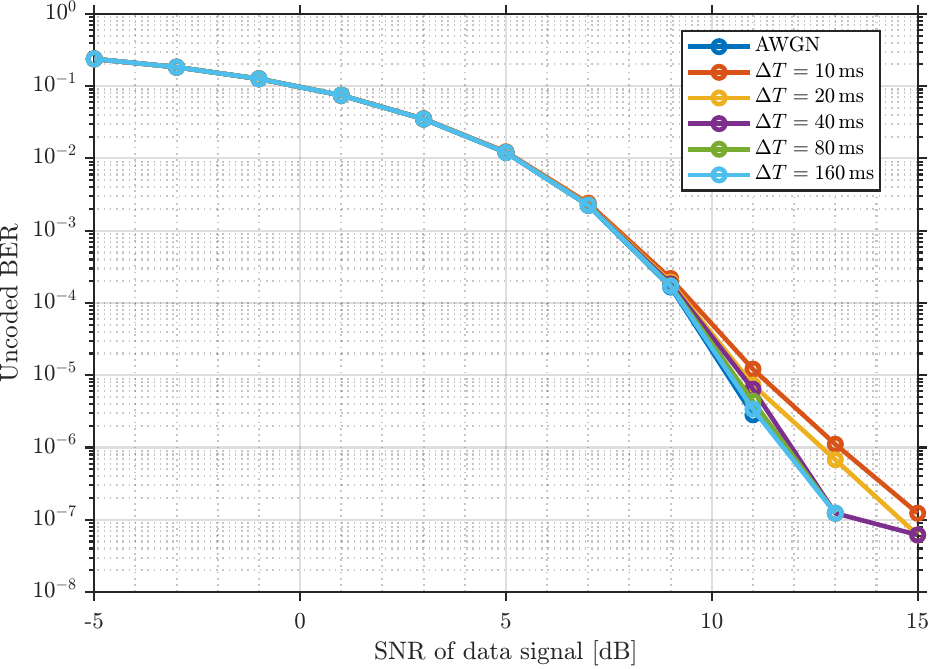}
    \caption{Uncoded \ac{BER} versus data-signal \ac{SNR} for different \ac{PRS} repetition periods $\Delta T=T_{\mathrm{rep}}$. The \ac{PRS} carrier power is 15~dB below that of the \ac{5G} data ($C_{\mathrm{data}}/I=15$~dB). \ac{FR1} parameters.}
    \label{fig:BER_15_FR1}
\end{figure}

\begin{figure}
    \centering
    \includegraphics[width=1\linewidth]{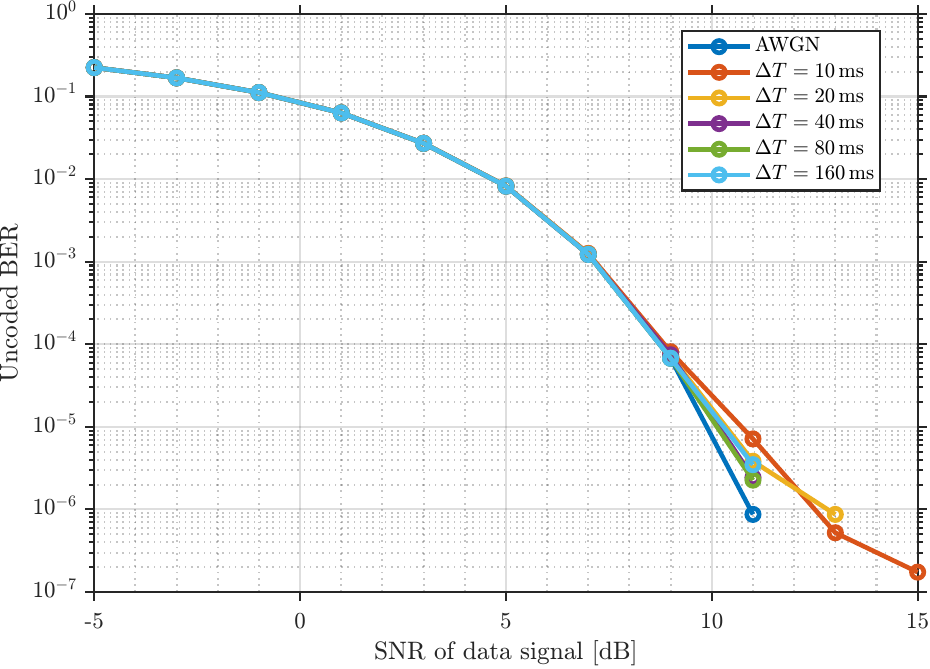}
    \caption{Uncoded \ac{BER} versus data-signal \ac{SNR} for different \ac{PRS} repetition periods $\Delta T=T_{\mathrm{rep}}$. The \ac{PRS} carrier power is 15~dB below that of the \ac{5G} data ($C_{\mathrm{data}}/I=15$~dB). \ac{FR2} parameters.}
    \label{fig:BER_15_FR2}
\end{figure}

Having established the acquisition regime, we now evaluate the impact of the \ac{PRS} on the data stream by measuring the uncoded \ac{BER}. The simulator transmits a continuous random \ac{QPSK} data stream that is periodically perturbed by \ac{PRS} occasions of duration $T_{\mathrm{PRS}}$, spaced by the repetition period $T_{\mathrm{rep}}$ (denoted $\Delta T$ in Figs.~\ref{fig:BER_20_FR1}--\ref{fig:BER_15_FR2}). The repetition period is configurable up to 10~s in \ac{3GPP}~\cite{3gpp_nr_2024}, but values larger than $\sim 160$~ms produce a behaviour visually indistinguishable from the \ac{AWGN} baseline at the considered \acp{SNR}. The horizontal axis of the \ac{BER} plots is the \ac{SINR} of the data signal at the receiver, defined as $\text{SINR}=C_{\mathrm{data}}/(N_0\,BW)$ in dB; the equivalent $C_{\mathrm{data}}/N_0$ is obtained by adding $10\log_{10}(BW)$ to the \ac{SNR}.

Figs.~\ref{fig:BER_20_FR1} and~\ref{fig:BER_20_FR2} report the \ac{BER} when the \ac{PRS} carrier-to-interference ratio is $C_{\mathrm{data}}/I=25$~dB, i.e., the \ac{PRS} carrier power received at the \ac{UE} is 25~dB below that of the serving-satellite data signal. This is the operating point that, when combined with $T_{\mathrm{PRS}}=10$ \ac{OFDM} symbols at \ac{FR2}, yields the high-confidence detection regime visible in Fig.~\ref{fig:ROC_FR2} (i.e., it corresponds to the $C/N_{0,\mathrm{eff}}$ value at which $P_d\to 1$ in that figure once the data signal of the serving satellite is added to the noise). At this $C_{\mathrm{data}}/I$, the \ac{BER} degradation is negligible for any tested $T_{\mathrm{rep}}$, confirming that a low-power \ac{PRS} is sufficient for the navigation function and benign for the data link.

Figs.~\ref{fig:BER_15_FR1} and~\ref{fig:BER_15_FR2} report an extreme case in which the \ac{PRS} carrier-to-interference ratio is 10~dB larger than in the previous case ($C_{\mathrm{data}}/I=15$~dB) in both \ac{FR1} and \ac{FR2}. Here the \ac{BER} degrades visibly, and the degradation is more pronounced at smaller $T_{\mathrm{rep}}$, i.e., at larger duty cycle $D=T_{\mathrm{PRS}}/T_{\mathrm{rep}}$ consistent with~\eqref{eq:BER_interference}, where the weight of the interfered-symbol term scales with $D$.

The assumed $C_{\mathrm{data}}/I$ values between the \ac{PRS} of a non-serving satellite and the data signal of the serving satellite are illustrative of the configurable power back-off available at the satellite \ac{HPA}; in a deployed system, the achieved $C_{\mathrm{data}}/I$ will also depend on the relative slant ranges and on the \ac{UE} antenna gain difference $g$, both of which can introduce several additional dB of asymmetry between the two received powers.

Finally, the \ac{FR1} and \ac{FR2} configurations show comparable \ac{BER} curves under the same $C_{\mathrm{data}}/I$ and $T_{\mathrm{rep}}$, indicating that the dominant scaling is set by the energy in the interfering \ac{PRS} burst.

\section{Conclusions}

This work has analysed the impact of a \ac{5G}-\ac{NR} \ac{PRS}-based positioning service hosted on the same downlink as the broadband data service in an \ac{NTN} scenario where the inter-satellite differential delay is much larger than the \ac{5G} slot duration. In such a scenario, periodic \ac{PRS} transmissions from non-serving satellites systematically collide with the data slots of the serving satellite. A closed-form, dual-shell statistical model of the absolute differential delay was derived and validated against an orbit simulation, and the resulting \ac{PRS} overlap probability per occasion was shown to be small.

The \ac{PRS} detection analysis shows that the navigation function can operate at a \ac{PRS} carrier power tens of dB below the data signal. The 6~dB shift between the \ac{FR1} and \ac{FR2} \acp{ROC} is consistent with the corresponding ratio of \ac{OFDM} symbol durations at the two \acp{SCS}, and is therefore an energy-of-integration effect. Operating at $C_{\mathrm{data}}/I\approx 25$~dB, the induced uncoded \ac{BER} degradation is negligible across a wide range of \ac{PRS} repetition periods and the \ac{BER} degradation scales with the duty cycle $D=T_{\mathrm{PRS}}/T_{\mathrm{rep}}$ as predicted by the analytical model.

Future work will address: tracking-loop robustness at the low-power \ac{PRS} regime identified here; coded-link performance with realistic \ac{3GPP} \ac{NR} \ac{LDPC} block codes, which interact with the burst nature of the interference differently from the uniform-\ac{AWGN} approximation used here; \ac{UE} antenna group-delay variation over the field of view and local oscillator instability and their impact on ranging accuracy, and system-level orchestration of beam scheduling so that users can periodically receive the \ac{PRS} from multiple sources without compromising the data service.

% \appendices
% \section*{APPENDIX. ACRONYMS}
% This paper uses an extensive number of acronyms, which are listed below to assist the reader:
  \acrodef{3G}{third generation}
  \acrodef{4G}{fourth generation}
  \acrodef{5G}{fifth generation}
  \acrodef{6G}{sixth generation}
  \acrodef{3GPP}{3rd generation partnership project}
  \acrodef{AI}{artificial intelligence}
  \acrodef{AMF}{access and mobility function}
  \acrodef{AoA}{angle of arrival}
  \acrodef{AoD}{angle of departure}
  \acrodef{AR}{augmented reality}
  \acrodef{ASIC}{application-specific integrated circuit}
  \acrodef{AWGN}{additive white Gaussian noise}
  \acrodef{BER}{bit error rate}
  \acrodef{BFN}{beamforming network}
  \acrodef{BO}{back-off}
  \acrodef{BW}{bandwidth}
  \acrodef{C/A}{civilian acquisition}
  \acrodef{CA}{cell average}
  \acrodef{CCDF}{complementary cumulative density function}
  \acrodef{CDF}{cumulative density function}
  \acrodef{CFAR}{constant false alarm rate}
  \acrodef{CFO}{carrier frequency offset}
  \acrodef{CMMB}{China mobile multimedia broadcasting}
  \acrodef{CP}{cyclix prefix}
  \acrodef{CRB}{commom resource block}
  \acrodef{CRLB}{Cramer-Rao lower bound}
  \acrodef{CSS}{chirp spread spectrum}
  \acrodef{CU}{centralized unit}
  \acrodef{DAC}{digital to analog converter}
  \acrodef{DBF}{digital beamforming}
  \acrodef{DFT}{discrete Fourier transform}
  \acrodef{DL-AoD}{downlink angle of departure}
  \acrodef{DLL}{delay locked loop}
  \acrodef{DL-OTDoA}{downlink observed time difference of arrival}
  \acrodef{DMRS}{demodulation reference signal}
  \acrodef{DSSS}{direct-sequence spread spectrum}
  \acrodef{DU}{distributed unit}
  \acrodef{DVB}{digital video broadcasting}
  \acrodef{E-CID}{enhanced cell id}
  \acrodef{ECDF}{empirical cumulative density function}
  \acrodef{E-SMLC}{evolved serving mobile location center}
  \acrodef{EIRP}{equivalent isotropic radiated power}
  \acrodef{EKF}{extended kalman filter}
  \acrodef{eNB}{evolved nodeb}
  \acrodef{EVM}{error vector magnitude}
  \acrodef{FHSS}{frequency hopping spread spectrum}
  \acrodef{FLL}{frequency locked loop}
  \acrodef{FFT}{fast Fourier transform}
  \acrodef{FFR}{full frequency re-use}
  \acrodef{FOV}{field of view}
  \acrodef{FSPL}{free space path loss}
  \acrodef{FR}{frequency region}
  \acrodef{FR1}{frequency region 1}
  \acrodef{FR2}{frequency region 2}
  \acrodef{FRF3}{frequency re-use factor 3}
  \acrodef{GDOP}{geometrical dilution of precision}
  \acrodef{GEV}{generalized extreme distribution}
  \acrodef{GPS}{global positioning system}
  \acrodef{gNB}{next generation base station}
  \acrodef{GNSS}{global navigation satellite system}
  \acrodef{GS}{ground station}
  \acrodef{HAPS}{high-altitude platform systems}
  \acrodef{HIL}{hardware-in-the-loop}
  \acrodef{HPA}{high power amplifier}
  \acrodef{IBO}{input back-off}
  \acrodef{ICAL}{integrated communications and localization}
  \acrodef{IDFT}{inverse discrete Fourier transform}
  \acrodef{IFFT}{inverse fast Fourier transform}
  \acrodef{IoT}{internet of things}
  \acrodef{IIoT}{industrial internet of things}
  \acrodef{ICI}{inter-carrier interference}
  \acrodef{IMU}{inertial measurement unit}
  \acrodef{ISI}{inter-symbol interference}
  \acrodef{JCAP}{joint communication and positioning}
  \acrodef{KPI}{key performance indicator}
  \acrodef{LBA}{link budget analysis}
  \acrodef{LBS}{location based services}
  \acrodef{LCS}{location-based services}
  \acrodef{LDPC}{low density parity check}
  \acrodef{LEO}{low earth orbit}
  \acrodef{LMC}{location management component}
  \acrodef{LMF}{location management function}
  \acrodef{LMMSE}{linear minimum mean squared error}
  \acrodef{LNA}{low noise amplifier}
  \acrodef{LoS}{line of sight}
  \acrodef{LPP}{localization positioning protocol}
  \acrodef{LPPa}{localization positioning protocol annex}
  \acrodef{LTE}{long term evolution}
  \acrodef{MCRB}{modified Cramer Rao bound}
  \acrodef{MEO}{medium Earth orbit}
  \acrodef{ML}{machine learning}
  \acrodef{Multi-RTT}{multi-cell round trip time}
  \acrodef{NF}{network function}
  \acrodef{NSGA-II}{non-dominated sorting genetic algorithm II}
  \acrodef{NGSO}{non geostationary satellite orbit}
  \acrodef{NLOS}{non-line of sight}
  \acrodef{NMSE}{normalized mean square error}
  \acrodef{NR}{new radio}
  \acrodef{NTN}{non-terrestrial network}
  \acrodef{OFDM}{orthogonal frequency-division multiplexing}
  \acrodef{OMP}{orthogonal matching pursuit}
  \acrodef{OTA}{over-the-air}
  \acrodef{OTDoA}{observed time differential of arrival}
  \acrodef{OTFS}{orthogonal time frequency space}
  \acrodef{PAPR}{peak-to-average power ratio}
  \acrodef{PDF}{probability density function}
  \acrodef{PDSCH}{physical downlink shared channel}
  \acrodef{PLL}{phase-locked loop}
  \acrodef{PNT}{positioning, navigation, and timing}
  \acrodef{POD}{precise orbit determination}
  \acrodef{PPP}{precise point positioning}
  \acrodef{PRN}{pseudo-random noise}
  \acrodef{PRS}{positioning reference signal}
  \acrodef{PSD}{power spectral density}
  \acrodef{PSS}{primary synchronization signal}
  \acrodef{PVT}{position, velocity and timing}
  \acrodef{QAM}{quadrature amplitude modulation}
  \acrodef{QoS}{quality of service}
  \acrodef{QPSK}{quadrature phase-shift keying}
  \acrodef{RAN}{radio access network}
  \acrodef{RAT}{radio-access-technology}
  \acrodef{RB}{resource block}
  \acrodef{RE}{resource element}
  \acrodef{RG}{resource grid}
  \acrodef{RedCap}{reduced capacity}
  \acrodef{RMSE}{root mean square error}
  \acrodef{ROC}{receiver operating characteristic}
  \acrodef{RTK}{real time kinematics}
  \acrodef{RRC}{radio resource control}
  \acrodef{SAN}{satellite access network}
  \acrodef{SBAS}{satellite based augmentation system}
  \acrodef{SCS}{subcarrier spacing}
  \acrodef{SDR}{software defined radio}
  \acrodef{SIB}{system information block}
  \acrodef{SIC}{sequential interference cancellation}
  \acrodef{SIR}{signal-to-interference ratio}
  \acrodef{SINR}{signal-to-interference plus noise ratio}
  \acrodef{SFN}{single frequency network}
  \acrodef{SLA}{service level agreement}
  \acrodef{SNR}{signal-to-noise ratio}
  \acrodef{SoO}{signal of opportunity}
  \acrodef{SoP}{signal of opportunity}
  \acrodef{SRS}{sounding reference signal}
  \acrodef{SRRC}{square root raised cosine}
  \acrodef{SSB}{synchronization signal block}
  \acrodef{SSC}{spectral separation coefficient}
  \acrodef{SSP}{subsatellite point}
  \acrodef{SSS}{secondary synchronization signal}
  \acrodef{TA}{timing advance}
  \acrodef{TC}{time coded}
  \acrodef{TC-OFDM}{time-coded orthogonal frequency division multiplexing}
  \acrodef{TDL}{tapped delay line}
  \acrodef{TDoA}{time difference of arrival}
  \acrodef{TN}{terrestrial network}
  \acrodef{ToA}{time of arrival}
  \acrodef{ToF}{time of flight}
  \acrodef{TS}{technical specification}
  \acrodef{TR}{technical report}
  \acrodef{TTFF}{time to first fix}
  \acrodef{UAV}{unmanned aerial vehicle}
  \acrodef{UE}{user equipment}
  \acrodef{UL-AoA}{uplink angle of arrival}
  \acrodef{UL-TDoA}{uplink time difference of arrival}
  \acrodef{UPA}{uniform planar array}
  \acrodef{VR}{virtual reality}
  \acrodef{WLAN}{wireless local area network}
  \acrodef{ZOH}{zero-order hold}

\bibliographystyle{IEEEtran}
\bibliography{references}

\end{document}